\documentstyle [12pt]{article}
\newcommand{\bea}{\begin{eqnarray}}
\newcommand{\eea}{\end{eqnarray}}
\newcommand{\be}{\begin{equation}}
\newcommand{\ee}{\end{equation}}

\begin{document}
\begin{center}
{\bf THE RIEMANN SURFACE OF A STATIC DISPERSION MODEL
AND REGGE TRAJECTORIES}\\
\vspace*{0.37truein}
V. A. Meshcheryakov \\
\vspace*{0.015truein}
{\it Joint Institute for Nuclear Research,\\
Bogoliubov Laborotory of Theoretica  Physics\\
Dubna 141 980, Moscow Region, Russia}
\end{center}
\begin{abstract}
The S-matrix in the static limit of a dispersion relation is a matrix of a
finite order N of meromorphic functions of energy $\omega$ in the plane with
cuts $(-\infty,-1],[+1,+\infty)$. In the elastic case it reduces to N
functions $S_{i}(\omega)$  connected by the crossing symmetry matrix A.  The
scattering of a neutral pseodoscalar meson with  an arbitrary angular momentum
l at a source with spin 1/2 is considered (N=2). The Regge trajectories of
this model are explicitly found.
\end{abstract}

The analytic structure of physical amplitudes in gauge theories with
confinement was investigated in ref.\cite{1}. It was shown that the analytic
structure of hadron physical amplitudes established in  old proofs of
dispersion relations remains valid in QCD. It is well known\cite{2}that
the static limit of a dispersion relation is equivalent to the system of
nonlinear integral eguations\cite{3}.Below,we will study this type of
equations reducing them to a nonlinear boundary value problem \cite{4}. It
consists of the following series of conditions on $S_i$, $S$--matrix elements:
\bea \label{} \left.  \begin{array}{l} a)~ S_i(z) - \mbox{meromorphic
functions in the complex plane} ~z~\\ ~~~~~~~~~~~~~~\mbox{with cuts}
 ~(-\infty,-1],[+1,+\infty), \\ b)~ S_i^*(z)=S_i(z^*), \\ c)~ \mid
 S_i(\omega+i0)\mid^2=1~\mbox{at}~ \omega\geq 1~~
 S_i(\omega+i0)=\lim\limits_{\epsilon\to+0}S_i(\omega+i\epsilon),\\ d)~
 S_i(-z)=\sum_{j=1}^{N}A_{ij}S_j(z).  \end{array}  \right.  \eea Real values
of the variable $z$ represent the total energy $\omega$  of a relativistic
particle scattered at a fixed center. The requirement for functions $S_i(z)$
being meromorphic results from the static limit of the scattering problem [5].
The elastic condition of unitarity (1c) is valid only on the right cut of the
plane $z$. On the left cut, functions $S_i(z)$ are given by the conditions of
crossing symmetry (1d). The matrix of crossing symmetry $A$ is defined by the group
under which the $S$-matrix is invariant; see, for instance \cite{4}.  Let us
write conditions (1) in the matrix form. To this end, we introduce the column
$~~S^{(0)}(z)=[S_1(z),~~S_2(z),\cdots,S_N(z)]$,~ where the upper index denotes
the physical sheet of the Riemann surface of the $S$-matrix.  Conditions
(1a,b,d) refer to the physical sheet, while the unitarity condition (1c) can
be extended to complex values of $\omega$, being of a component-wise form,
$S_i^{(0)}(z)S_i^{(1)}(z)=1$. The matrix form of the unitarity condition (1c)
is derived by the nonlinear operation of inversion $I$ according to the
formula $~~IS(z)=[1/S_1(z),1/S_2(z),\cdots,1/S_N(z)].~$ As a result,
conditions (1a,b,c,d)  assume the form \bea \label{} \left.  \begin{array}{l}
a)~ S^{(0)}(z) - \mbox{a column of meromorphic functions in the complex plane}~
z~\\ ~~~~~~~~~~~~~~~\mbox{with cuts}~~ (-\infty,-1],[+1,+\infty)\\ b)~
{S^{(0)}}^*(z)=S^{(0)}(z^*)\\ c)~ S^{(1)}(z)=IS^{(0)}(z) \\ d)~
S^{(0)}(-z)=AS^{(0)}(z) \end{array} \right.  \eea Analytic continuation onto
unphysical sheets will be defined as follows:  \be
S^{(p)}(z)=(IA)^pS^{(0)}(z(-1)^p).  \ee By using the definition (3), we can
easily continue the unitarity condition (2c) and crossing symmetry (2d) on to
unphysical sheets \be IS^{(p)}(z)=S^{(1-p)}(z), AS^{(p)}(z)=S^{(-p)}(-z) \ee
and we arrive at the formula \be (IA)^qS^{(p)}(z)=S^{(q+p)}(z(-1)^q).  \ee For
example, the scattering of a neutral pseudoscalar pions at a fixed nucleon
with spin $1/2$ is defined by the condition (1) and the two-row matrix \be
\label{matrix}
A=\frac{1}{2l+1}\left(\begin{array}{rc}-1&2l+2\\2l&1\end{array}\right),~~l\in N.
\ee
Let us introduce the function $X= S_{1}/S_{2}$ and consider it for $z=0$.
Then the continuation of X on to the first unphysical sheet is determined by
the rule
$$
  X^{(1)}=\frac{2lX^{(0)}+1}{-X^{(0)}+(2l+2)}
$$
and together with the crossing symmetry condition (4) gives
the following expression for $X^{(n)}$
 \be X^{(n)}=\frac{n-(l+1)}{n+l},~~X^{(0)}=-(1+1/l)~. \ee
Thus, on any unphysical sheet n the ratio $S_{1}/S_{2}$ is defned at $z=0$
and for construction of $S_{1}$and$S_2$ it is sufficient to find any of them.
Let us denote $S_2$ by $\varphi=S_2$. This function is determind by the
system of functional eguations
\be \varphi^{(n)}\varphi^{(1-n)}=1~, \ee
\be
\frac{\varphi^{(n)}}{\varphi^{(-n)}}=\frac{n+l}{n-l}~,
\ee
which follows from the unitarity and the crossing symmetry conditions (4) on
the unphysical sheets. Here  only those equalities are used from (4),which
were not used for derivation of eq.(7).  Equation (8) has an obvious
solution in the ring of meromorphic functions \be
\varphi^{(n)}=\frac{G(n)}{G(1-n)}~, \ee where $G(n)$ is an entire function.
 Solution (10) can be represented in  another form
$\ln\varphi^{(n)}=g(n-1/2)~,$ where $g(n-1/2)$ is any odd function of its
argument. That form of $\ln\varphi^{(n)}$ is convenient for the solution to
eq.(9) which is now of the form $$ g(n+1)+g(n)=\ln\frac{n+1/2+l}{n+1/2-l}~.
$$
Ae partial solution of this nonhomogeneous difference equation  can be
found by subsequent  substitutions of a unknown functions according to the
formulae
$$g_{m}(n)=g_{m+1}(n)+\ln\frac{n+(-1)^{m}\alpha_{m+1}}{n-(-1)^{m}\alpha_{m+1}}~,$$
where $\alpha_{k}=1/2+l-k$ and $g_{0}(n)=g(n)$.
The function $g_k$ obeys the equation
$$g_{k}(n+1)+g_{k}(n)=\ln\frac{n+1/2+(-1)^{k}(l-k)}{n+1/2-(-1)^{k}(l-k)}.$$
It is clear that
\be
 g_{l}(n+1)+g_{l}(n)=0
\ee
 and a general solution to this equation gives a trivial solution of the
 problem (1) which does not depend on l.Therefore, one gets\cite{5}
\be
\varphi^{(n)}=\prod\nolimits_{m=1}^{l}\frac{n-1/2-(-1)^{m}(1/2+l-m)}{n-1/2+(-1)^{m}(1/2+l-m)}~.
\ee
One has an infinite product in formulae (12) for noninteger$l \in R$.  Now eq.(11) is of the form \be g(n+1)+g(n)=ln(-1) \ee In this case
one has instead of eq.(12) \be
\varphi^{(n)}=\psi(n)\frac{\Gamma[-\frac{n+l}{2}+1]\Gamma[\frac{n-l}{2}]}
   {\Gamma[-\frac{n-1-l}{2}+1]\Gamma[-\frac{n-1+l}{2}]}
\ee
where $\psi(n)$ is a general solution of eq.(13) with properties
\be
\psi(n+1)\psi(n)=1,~~~\psi(n)\psi(-n)=1
\ee
Till now one of the unitarity conditions (1c) was not used and it dives the
following result
\be
n(z)=1/\pi\arcsin z+i\sqrt{z^2-1}\beta(z),
\ee
where $\beta(z)=-\beta(-z)$ --is a meromorphic function.
Equation(16) shows that the Riemann surface of the model has an algebraic
branch points at $z=\pm1$ and a logarithmic one at infinify. Now  formulae
(7,14,15,16) give the general solution to the problem (1) for matrix (6). The
function $\psi$ can be determined from the requirement that eq.(14) turns
to to eq.(12) for integer l.  This gives $\psi(n)=-\cot (n)$ for l even
and $\psi(n)=-\tan (n)$ for l odd.\\ Let us remind that in eq.(14) $l\in R$,
but it is clear that this relation can be continue to $l\in C$ and allous
 explicsitl determinaion of  the Regge tragectories with definite signature
$l^{\pm}_{k}(z)$.  The common part of the set  of Regge tragectories set for
$J_{\pm}=l\pm 1/2$ is of the form $l^{\pm}(z)=\{2-n(z)+2k,n(z)+2k~|~
k=0,1,2\cdots\}$.  The Regge trajectories for $J_{-}=l-1/2$ contained one
additional trajectory $l^{\pm}_{J_{-}}(z)=-n(z)$.  All the Regge trajectories
of the model depends on one function $\beta (z)$.

\end{document}